%% file: main.tex
\documentclass[conference]{IEEEtran}
\IEEEoverridecommandlockouts
\usepackage{cite}
\usepackage{amsmath,amssymb,amsfonts}
\usepackage{algorithmic}
\usepackage{algorithm}
\usepackage{graphicx}
\usepackage{textcomp}
\usepackage{xcolor}
\usepackage{comment}
\usepackage{orcidlink}
\usepackage{booktabs}
\usepackage{multirow}
\usepackage{url}
\usepackage{tikz}
\usetikzlibrary{arrows.meta,positioning,calc,fit}
\hypersetup{hidelinks}

\newcommand{\sys}{PlaneZip}
\newcommand{\para}[1]{\smallskip\noindent\textbf{#1}}

\def\BibTeX{{\rm B\kern-.05em{\sc i\kern-.025em b}\kern-.08em
    T\kern-.1667em\lower.7ex\hbox{E}\kern-.125emX}}
\begin{document}
\bstctlcite{IEEEexample:BSTcontrol}

\title{PlaneZip: Lossless Bit-Plane Compression of Training Activations Offloaded over PCIe}

\author{\IEEEauthorblockN{
Zhufan Gu\IEEEauthorrefmark{1},
Qizhe Tang\IEEEauthorrefmark{2},
Yuxin Wang\IEEEauthorrefmark{3},
Xun Zhou\IEEEauthorrefmark{4},
Hui Feng\IEEEauthorrefmark{5}
}\\
  \IEEEauthorblockA{
  \IEEEauthorrefmark{1}Trine University
  \quad
  \IEEEauthorrefmark{1}Santa Clara University
  \quad
  \IEEEauthorrefmark{3}UT Austin
  \quad
  \IEEEauthorrefmark{4}Independent Researcher
  \quad
  \IEEEauthorrefmark{5}UC Merced
  }
}

\author{\IEEEauthorblockN{
Zhufan Gu,
Qizhe Tang,
Yuxin Wang,
Xun Zhou,
Hui Feng
}}

\maketitle

\input{sections/abstract}
\input{sections/intro}
\input{sections/background}
\input{sections/design}
\input{sections/method}
\input{sections/eval}
\input{sections/related}
\input{sections/conclusion}

\bibliographystyle{IEEEtran}
\bibliography{references}

\end{document}

%% file: sections/abstract.tex
\begin{abstract}
Training under a device-memory budget relies on tensor rematerialization: an evicted activation is either swapped to host memory or dropped and recomputed. Because GPU compute has outgrown PCIe bandwidth, cost-model planners such as MegTaiChi, which charge a swap its full transfer time, almost never swap, and swaps that fill idle link time carry only a small fraction of the saved activations. Saved activations, however, are homogeneously typed arrays with bit-level redundancy beyond zeros, which lightweight bit-plane compressors can remove at line rate. We present PlaneZip, which compresses swapped tensors on the copy path and plans around the compression. Its codec picks, for every 128-byte block, the smallest of zero-value, delta bit-plane, bit-plane, and a new zero-masked bit-plane encoding. A per-tensor predictor reuses the ratio of the previous iteration, and the planner budgets link time in compressed transfer times. On activations captured from five models trained on real data, the codec compresses BF16 activations by 1.71x and FP32 activations by 1.39x (geometric means), against 1.18x and 1.20x for zero-value compression. In trace-driven simulation of a PCIe 3.0 consumer GPU and a PCIe 4.0 datacenter GPU at half the activation memory, PlaneZip cuts the rematerialization overhead by 47\% and 31\% relative to a bandwidth-aware planner without compression, and by 73\% and 54\% relative to MegTaiChi's cost model, with an estimated engine area of 0.036 square millimeters.
\end{abstract}


%% file: sections/intro.tex
\section{Introduction}
\label{sec:intro}

The activations that the forward pass saves for the backward pass dominate the memory footprint of DNN training~\cite{chen2016sublinear,korthikanti2023reducing}. When they exceed the device memory, systems either oversubscribe it through unified virtual memory (UVM), which migrates pages on demand~\cite{lin2025understanding,lin2025forest}, or rematerialize the activations explicitly. A saved tensor is either swapped to host memory and prefetched before its backward use~\cite{rhu2016vdnn,huang2020swapadvisor,peng2020capuchin}, or dropped and recomputed from its inputs~\cite{chen2016sublinear,jain2020checkmate,kirisame2021dtr}. MegTaiChi~\cite{hu2022megtaichi} coordinates the two: when memory runs out, it evicts the tensor with the lowest cost, charging a swap its transfer time and a drop its recomputation time.

The two mechanisms scale differently. Recomputation runs at GPU speed, and the two GPUs we model differ by 23$\times$ in peak math throughput, from a Turing-class consumer card training in FP32 to an Ampere-class datacenter card training in BF16, while their host links differ by only 2$\times$ (PCIe~3.0 and 4.0). For the median saved byte, a swap takes 10--13$\times$ longer than recomputing the same tensor (Section~\ref{sec:motiv}), so a per-tensor cost model almost never picks it: in our simulations MegTaiChi's policy reduces to pure recomputation in 58 of 70 configurations. Swapping still pays when it uses link time that would otherwise idle, which bandwidth-aware planners such as Capuchin~\cite{peng2020capuchin} exploit. That idle time is short, though: during one iteration the link can move only 6--37\% of the saved bytes of the five models we study.

Compression enlarges that budget. Saved activations are large arrays of a single type, the property that Bit-Plane Compression (BPC)~\cite{kim2016bpc} exploits in GPU memory traffic. They contain zeros produced by ReLU and by causal attention masks, and neighboring floating-point values tend to share their high-order exponent bits. Hardware block compressors designed for GPU and main memory~\cite{kim2016bpc,choukse2018compresso} encode a 128-byte block in a few cycles, fast enough to sit on the copy path. The compressing DMA engine cDMA~\cite{rhu2018cdma} applied zero-value compression (ZVC) to swapped activations, which captures ReLU sparsity but nothing else, and its swap schedule did not change with the compression it achieved.

This paper presents \sys{}, which compresses swapped tensors on the copy path and plans around the compression. It rests on three observations from activations that we capture in real training iterations:
\begin{itemize}
\item \textbf{Activations compress beyond their zeros.} A per-block choice among four lightweight encodings, including a zero-masked bit-plane encoding that we add, compresses BF16 activations by 1.71$\times$ (geometric mean over five models), against 1.18$\times$ for ZVC (Section~\ref{sec:eval-cr}).
\item \textbf{Compressibility is a stable per-tensor property.} Training repeats the same operator sequence every iteration, which MegTaiChi exploits for its memory plans. Predicting each tensor's compressed size from the previous iteration errs by less than 1\% of the transferred bytes (Section~\ref{sec:eval-pred}).
\item \textbf{Only a compression-aware planner turns compression into speed.} Compressing the swaps chosen by a compression-unaware planner shortens transfers but leaves the freed link time unused. Budgeting the link with compressed transfer times, and ranking tensors by the recomputation or memory they save per second of transfer, recovers it (Section~\ref{sec:eval-why}).
\end{itemize}

We make the following contributions:
\begin{itemize}
\item A characterization of the compressibility of saved activations in three CNNs and two Transformers, for six codecs in FP32 and BF16.
\item The \sys{} codec: element-width-aware selection among ZVC, BPC's delta bit-plane transform, a bit-plane transform without the delta, and a new zero-masked bit-plane encoding, packed into 4\,KB chunks that decode in parallel, with a bit-exact reference implementation.
\item A per-tensor compressibility predictor and a compression-aware swap and recomputation planner.
\item A trace-driven evaluation on a PCIe~3.0 consumer GPU and a PCIe~4.0 datacenter GPU, in which \sys{} reduces the rematerialization overhead at 50\% of the activation memory by 47\% and 31\% relative to a bandwidth-aware planner without compression.
\end{itemize}

%% file: sections/background.tex
\section{Background and Motivation}
\label{sec:background}

\subsection{Tensor Rematerialization}
A training iteration runs a forward pass that saves activations, a backward pass that consumes them in reverse order, and an optimizer step. A saved tensor $t$ of $m_t$ bytes is idle between its last forward use and its first backward use, and the peak memory occurs near the boundary of the two passes, when nearly every saved tensor is alive. Rematerialization frees $t$ during its idle interval in one of two ways:
\begin{itemize}
\item \textbf{Swap.} Copy $t$ to pinned host memory once it is final, free it when the copy completes, and prefetch it before its first backward use~\cite{rhu2016vdnn,huang2020swapadvisor}. A swap costs nothing if both copies overlap with computation, and stalls the GPU otherwise.
\item \textbf{Drop.} Free $t$ after its last forward use and replay the operators that produced it before its backward use~\cite{chen2016sublinear,jain2020checkmate}. The replay costs $C_r(t)$, which grows if the inputs of those operators were dropped as well~\cite{kirisame2021dtr}.
\end{itemize}

MegTaiChi's dynamic tensor evicting~\cite{hu2022megtaichi} decides between the two when memory runs out. It evicts the resident tensor that minimizes
\begin{equation}
\frac{\min\left(C_r(t),\, C_s(t)\right)}{m_t \cdot s_t}, \qquad C_s(t) = \frac{m_t}{\gamma},
\label{eq:dte}
\end{equation}
where $s_t$ is the staleness of $t$ and $\gamma$ the host-link bandwidth, and it swaps $t$ only if $C_s(t) < C_r(t)$. (The full cost also accounts for memory fragmentation and for how often $t$ has been recomputed.) Capuchin~\cite{peng2020capuchin} instead charges a swap only the transfer time that it cannot hide behind computation. UVM moves the same data at page granularity without a plan, so its cost depends on access-aware eviction and prefetching~\cite{lin2025forest,lin2026autouvm}.

\subsection{Memory Compression}
Hardware compressors for caches and memory operate on small blocks and must decode in a few cycles~\cite{pekhimenko2012bdi}. BPC~\cite{kim2016bpc} targets the homogeneously typed arrays that dominate GPU memory. Its Delta-BitPlane-XOR (DBX) transform subtracts neighboring 32-bit words of a 128-byte block, transposes the deltas into bit-planes, and XORs adjacent planes. Small deltas leave long runs of all-zero planes, which a simple prefix code compresses. Compresso~\cite{choukse2018compresso} adapted BPC to compress main memory transparently to the operating system. It found that the DBX transform is not always beneficial and compresses each block with and without it, and it predicts incompressible pages to avoid the data movement that compression otherwise causes.

Swapping turns the GPU and the host into a two-tier memory: a small fast tier and a large tier behind a narrow link. Heterogeneous memory systems, from hybrid main memory to CPU-FPGA SoCs~\cite{du2022comprehensive}, place data across such tiers, sometimes compressing the fast tier to raise its capacity~\cite{li2023baryon}. Yokoyama et al.~\cite{yokoyama2026rvv} found that splitting a PC-indexed reuse predictor~\cite{wu2011ship} by access type (vector or scalar) improves its accuracy at equal capacity. We apply the same idea to compressibility: a signature that identifies each tensor by the operator that produced it predicts its compressed size (Section~\ref{sec:pred}).

\subsection{Motivation}
\label{sec:motiv}
\begin{figure}[t]
\centering
\includegraphics[width=\columnwidth]{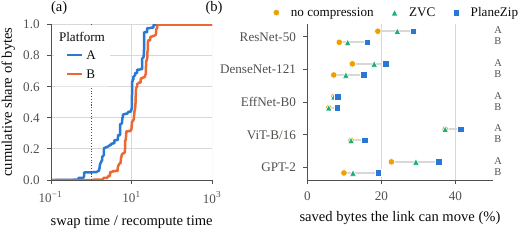}
\caption{(a) Swap-to-recompute cost ratio of every saved tensor of the five models, weighted by size. (b) Share of the saved bytes that the host link can move during one iteration on platforms A and B, without compression, with ZVC, and with the \sys{} codec.}
\label{fig:motiv}
\end{figure}

Figure~\ref{fig:motiv} quantifies the gap with the traces and platforms of Section~\ref{sec:method}. Figure~\ref{fig:motiv}a plots, for every saved tensor of the five models, the ratio between its transfer time $C_s$ and the time to recompute it with its inputs resident. On the PCIe~3.0 consumer platform (A), 95\% of the saved bytes cost more to swap than to recompute, and the median ratio is 10.1$\times$; on the PCIe~4.0 datacenter platform (B), over 99.9\% do, and the median is 12.6$\times$. Equation~\ref{eq:dte} therefore picks recomputation for nearly every tensor. Figure~\ref{fig:motiv}b shows how much swapping can still contribute: during one iteration, the link can move 7--37\% of the saved bytes on platform A and 6--12\% on platform B. Compressing the transfers with ZVC raises these shares by 1.2$\times$ on average, and the \sys{} codec by 1.4$\times$ (A) and 1.7$\times$ (B).

%% file: sections/design.tex
\section{\sys{} Design}
\label{sec:design}

\begin{figure}[t]
\centering
\resizebox{\columnwidth}{!}{\input{figs/overview.tex}}
\caption{\sys{} overview. The engine encodes each 128\,B block in four ways in parallel and keeps the smallest output or the raw block.}
\label{fig:overview}
\end{figure}
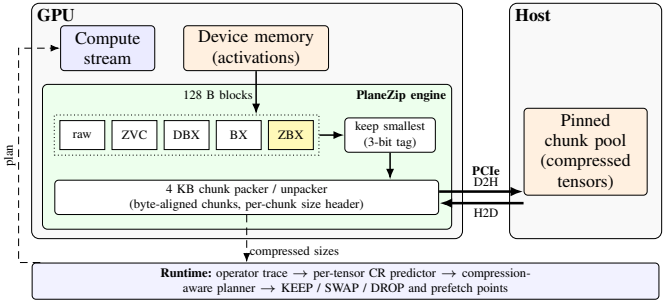

Figure~\ref{fig:overview} shows the three parts of \sys{}. A compression engine on the GPU's copy path compresses every swapped tensor on its way to host memory and decompresses it on the way back. A runtime predicts the compressed size of each saved tensor from the sizes that the engine reported in earlier iterations. A planner uses these sizes to decide, for every saved tensor, whether to keep, swap, or drop it, and when to prefetch it.

\subsection{Codec}
\label{sec:codec}
The engine splits a tensor into 128-byte blocks, the granularity of BPC~\cite{kim2016bpc}, and interprets each block as $N$ symbols of $w$ bits, where $w$ is the element width of the tensor: 64 symbols for BF16 and 32 for FP32. The runtime passes the element type in the copy descriptor; other types use $w=32$. Each block is encoded in four ways:
\begin{itemize}
\item \textbf{ZVC}~\cite{rhu2018cdma}: an $N$-bit mask of non-zero symbols followed by the non-zero symbols.
\item \textbf{DBX}: BPC's transform. A base symbol, then the $w{+}1$ bit-planes of the $N{-}1$ deltas between neighboring symbols, with adjacent planes XORed.
\item \textbf{BX}: the $w$ bit-planes of the symbols themselves, with adjacent planes XORed and no delta. Neighboring floating-point values often share their high-order exponent bits but not their low-order bits, so their deltas can be large while the high bit-planes stay constant. Compresso likewise found that skipping BPC's transform sometimes compresses better~\cite{choukse2018compresso}.
\item \textbf{ZBX}: an $N$-bit zero mask, then BX over the $n$ non-zero symbols compacted to the front of the block, so each plane is $n$ bits wide. In a ReLU output or a causally masked attention matrix, zeros are interleaved with positive values of similar magnitude. ZVC stores the survivors verbatim, and in the planes of BX or DBX the zeros break every run; ZBX removes the zeros first and then compresses the shared high-order bits of the survivors.
\end{itemize}
DBX, BX, and ZBX code each plane with BPC's prefix code~\cite{kim2016bpc}: 3 bits for an all-zero plane and 6 or 7 for a run of them, 5 bits for an all-ones plane or one whose pre-XOR plane is zero, 5 bits plus the position for one or two adjacent ones, and a 1-bit flag before any other plane, stored verbatim. The field widths scale with the number and width of the planes. The engine keeps the smallest of the four encodings and the raw block and spends a 3-bit tag on the choice.

\para{Stream format.} Thirty-two consecutive blocks form a 4\,KB chunk. A chunk is byte-aligned and preceded by a 16-bit compressed size, so the decompressor finds every chunk with a prefix sum over the headers and decodes chunks in parallel; a chunk that would grow is stored raw. On the host, compressed tensors occupy fixed-size chunks of a pinned pool, so their variable sizes do not fragment the allocator, the same reason Compresso allocates compressed pages in fixed 512\,B chunks~\cite{choukse2018compresso}. Every ratio we report includes tags, headers, and padding.

\para{Validation.} We implemented every encoder twice: a vectorized size model, used for all measurements, and a bit-exact encoder and decoder. On 4{,}000 blocks sampled from real FP32 and BF16 activations of the five models and 1{,}800 synthetic corner cases, the decoder reproduces every block and the bit counts of the two implementations agree exactly.

\subsection{Compressibility Predictor}
\label{sec:pred}
The planner needs the compressed size of a tensor before the tensor exists. Training repeats the same operator sequence every iteration, so \sys{} identifies a saved tensor by the position of its producing operator in that sequence and its output slot, the tensor-level analogue of the PC signature of a reuse predictor~\cite{wu2011ship}. The runtime stores one compression ratio per signature, a table of a few hundred entries. The engine reports the compressed size of every swap, the sum of its chunk headers, and the runtime overwrites the entry. Tensors that the plan does not swap are measured in a profiling iteration, in which the engine compresses them into a scratch buffer and discards the output. Until a signature has a measurement, the planner assumes a ratio of one.

\subsection{Compression-Aware Planner}
\label{sec:planner}
The planner runs on the host whenever the operator trace or the predicted ratios change, which in steady-state training happens only during profiling. Algorithm~\ref{alg:plan} lists it. A tensor $t$ with predicted ratio $c_t$ occupies a copy queue for
\begin{equation}
x_t = \min\!\left(\frac{m_t}{\gamma},\ \max\!\left(\frac{m_t}{c_t\,\gamma},\ \frac{m_t}{B_{\mathit{CE}}}\right)\right),
\label{eq:xfer}
\end{equation}
where $B_{\mathit{CE}}$ is the engine throughput: the transfer is limited by the link or by the engine, and the runtime bypasses the engine when compression would not shorten it.

The first stage fills idle link time with swaps, and compression enters it twice. The stage admits swaps until their total copy time reaches its budget, so shorter compressed transfers admit more of them; and it ranks candidates by what they buy per second of copy time: the recomputation time a swap avoids, $C_r(t)/x_t$, or the memory it frees, $m_t/x_t$. Both rankings favor compressible tensors. A candidate is admitted only if both copy queues still meet their deadlines: the device-to-host queue, ordered by production time, must drain before the forward pass ends, and the host-to-device queue, ordered by first backward use, must deliver each tensor before its use. The second stage covers any remaining deficit with recomputation, using the DTR heuristic $C_r(t)/(m_t \cdot \mathit{idle}_t)$~\cite{kirisame2021dtr}, where $C_r(t)$ includes the replay of dropped inputs and is charged again for every dropped tensor whose replay needs $t$. It skips any candidate whose replay would push memory above the budget. Finally, each swapped tensor is prefetched at the earliest point from which the budget holds until its first use.

The stage-one budget is a fraction $\eta$ of the shorter of the forward and backward compute times. Which ranking and budget work best depends on the model, so the planner builds a small portfolio of plans: the two rankings above and recomputation saved per byte, $C_r(t)/m_t$, each with $\eta \in \{1, 0.75, 0.5, 0.25\}$, plus plans built with uncompressed sizes, a plan from the per-tensor cost model of Equation~\ref{eq:dte} with compressed transfer times, and a recomputation-only plan. It simulates each of these 18 plans with the predicted ratios (Section~\ref{sec:method}), which takes milliseconds per plan, and keeps the fastest. Under its predictions, it therefore never does worse than recomputation alone, than compressing the swaps of an uncompressed plan, or than the compressed cost model.

\begin{algorithm}[t]
\caption{One candidate plan for budget $M$}
\label{alg:plan}
\begin{algorithmic}[1]
\small
\REQUIRE saved tensors $T$, predicted ratios $c_t$, budget $M$, fraction $\eta$, ranking $\rho(t) \in \{C_r(t)/x_t,\ m_t/x_t,\ C_r(t)/m_t\}$
\STATE $P \leftarrow$ KEEP for all $t$; \ $B \leftarrow \eta \cdot \min(T_{\mathit{fwd}}, T_{\mathit{bwd}})$
\FORALL{$t \in T$ in decreasing $\rho(t)$, with $x_t$ from Eq.~\ref{eq:xfer}}
  \IF{$x_t$ fits in $t$'s windows, $\sum_{\mathrm{SWAP}} x + x_t \le B$, and both queues meet their deadlines}
    \STATE $P[t] \leftarrow$ SWAP
  \ENDIF
  \STATE \textbf{if} peak memory of $P \le M$ \textbf{then break}
\ENDFOR
\WHILE{peak memory of $P > M$}
  \STATE $t \leftarrow \arg\min\, C_r(t)/(m_t \cdot \mathit{idle}_t)$ over KEEP tensors idle at the peak
  \STATE $P[t] \leftarrow$ DROP
\ENDWHILE
\STATE schedule each prefetch at the earliest point where memory stays $\le M$
\RETURN $P$
\end{algorithmic}
\end{algorithm}

\subsection{Hardware Cost}
\label{sec:hw}
BPC's compressor occupies 48{,}000\,$\mu$m$^2$ in a 40\,nm process at 800\,MHz and encodes 128 bytes every four cycles with a latency of 11 cycles~\cite{kim2016bpc}. A \sys{} lane adds the BX and ZBX encoders, which share BPC's bit-plane transposition and pattern logic, a 64-entry compaction network for ZBX, and the size comparison. We estimate a lane at three times the area of BPC's compressor and assume decompressors of equal size. Two compressor lanes for device-to-host copies and two decompressor lanes for host-to-device copies, each at 1.5\,GHz, deliver 96\,GB/s per direction, above the 64\,GB/s that we assume in the evaluation. Scaling 40\,nm logic to a 7\,nm process by a conservative 16$\times$ in density gives 0.036\,mm$^2$ for the four lanes, 0.004\% of an 826\,mm$^2$ A100 die~\cite{nvidia2020a100}; BPC's 7\,mW of active power at 40\,nm~\cite{choukse2018compresso} suggests well under 0.1\,W. The added latency, tens of cycles per block, is negligible next to the microseconds of a PCIe transfer.

%% file: figs/overview.tex
\begin{tikzpicture}[
  font=\scriptsize,
  box/.style={draw, rounded corners=1.5pt, align=center, inner sep=1.5pt},
  enc/.style={draw, fill=white, align=center, inner sep=1pt, minimum width=0.62cm, minimum height=0.40cm, font=\tiny},
  arr/.style={-{Latex[length=1.5mm]}, semithick},
  big/.style={-{Latex[length=1.8mm]}, line width=1.0pt},
  darr/.style={-{Latex[length=1.3mm]}, densely dashed},
  lbl/.style={font=\tiny, inner sep=1pt, align=center}
]
\draw[rounded corners=3pt, fill=gray!5] (0,0) rectangle (5.95,-3.25);
\node[anchor=north west, font=\scriptsize\bfseries, inner sep=2pt] at (0,0) {GPU};
\draw[rounded corners=3pt, fill=gray!5] (6.6,0) rectangle (8.7,-3.25);
\node[anchor=north west, font=\scriptsize\bfseries, inner sep=2pt] at (6.6,0) {Host};
\node[box, fill=blue!8, minimum width=1.3cm, minimum height=0.62cm] (sm) at (1.05,-0.62) {Compute\\stream};
\node[box, fill=orange!12, minimum width=2.0cm, minimum height=0.62cm] (hbm) at (3.10,-0.62) {Device memory\\(activations)};
\draw[rounded corners=2pt, fill=green!7] (0.15,-1.12) rectangle (5.8,-3.12);
\node[anchor=north east, font=\tiny\bfseries, inner sep=2pt] at (5.8,-1.12) {PlaneZip engine};
\node[enc] (e0) at (0.70,-1.82) {raw};
\node[enc] (e1) at (1.42,-1.82) {ZVC};
\node[enc] (e2) at (2.14,-1.82) {DBX};
\node[enc] (e3) at (2.86,-1.82) {BX};
\node[enc, fill=yellow!35] (e4) at (3.58,-1.82) {ZBX};
\node[draw, densely dotted, inner sep=2pt, fit=(e0)(e4)] (grp) {};
\node[box, fill=white, minimum width=1.25cm, minimum height=0.40cm, font=\tiny] (sel) at (4.95,-1.82) {keep smallest\\(3-bit tag)};
\draw[arr] (grp.east) -- (sel.west);
\node[box, fill=white, minimum width=5.3cm, minimum height=0.40cm, font=\tiny] (pk) at (2.97,-2.68)
  {4 KB chunk packer / unpacker\\(byte-aligned chunks, per-chunk size header)};
\draw[arr] (sel.south) -- (sel.south|-pk.north);
\draw[arr] (hbm.south) -- node[lbl, left, pos=0.55] {128 B blocks} (hbm.south|-grp.north);
\node[box, fill=orange!12, minimum width=1.7cm, minimum height=1.2cm] (pool) at (7.65,-2.05) {Pinned\\chunk pool\\(compressed\\tensors)};
\draw[big] ([yshift=0.08cm]pk.east) -- ([yshift=0.08cm]pk.east-|pool.west);
\draw[big] ([yshift=-0.08cm]pk.east-|pool.west) -- ([yshift=-0.08cm]pk.east);
\node[lbl, font=\tiny\bfseries, anchor=south] at (6.275,-2.41) {PCIe};
\node[lbl, anchor=south] at (6.275,-2.565) {D2H};
\node[lbl, anchor=north] at (6.275,-2.79) {H2D};
\node[box, fill=blue!5, minimum height=0.5cm, anchor=north west, text width=8.55cm, font=\tiny] (rt) at (0,-3.58)
  {\textbf{Runtime:} operator trace \(\rightarrow\) per-tensor CR predictor \(\rightarrow\) compression-aware planner \(\rightarrow\) KEEP / SWAP / DROP and prefetch points};
\draw[darr] (2.97,-2.92) -- node[lbl, right, pos=0.78] {compressed sizes} (2.97,-3.58);
\draw[darr] (0.12,-3.58) -- (-0.18,-3.58) -- node[lbl, sloped, above, pos=0.5] {plan} (-0.18,-0.62) -- (sm.west);
\end{tikzpicture}

%% file: sections/method.tex
\section{Methodology}
\label{sec:method}

\para{Workloads.} Table~\ref{tab:workloads} lists the five models. The three CNNs and ViT-B/16 train on the 1{,}000 ImageNet sample images (one per class)~\cite{russakovsky2015imagenet} with the standard random-resized-crop and flip augmentation, starting from public timm checkpoints~\cite{wightman2019timm,wightman2021rsb} that represent steady-state training. GPT-2 (124M)~\cite{radford2019gpt2}, in the Hugging Face implementation~\cite{wolf2020transformers}, trains from scratch on WikiText-2~\cite{merity2017wikitext} with a 50{,}257-token byte-level BPE vocabulary learned on the same corpus. We train it for 300 steps, which lowers the validation loss from 10.9 to 6.0, so that its activations come from a model in training rather than from random weights.

\begin{table}[t]
\centering
\caption{Workloads. Saved activations at the training batch size in FP32.}
\label{tab:workloads}
\setlength{\tabcolsep}{3.2pt}
\footnotesize
\begin{tabular}{lrrrrr}
\toprule
Model & Params & Batch & Saved & Saved & Ops \\
 & & & tensors & GB & \\
\midrule
ResNet-50~\cite{he2016resnet} & 25.6M & 128 & 216 & 11.0 & 896 \\
DenseNet-121~\cite{huang2017densenet} & 8.0M & 128 & 494 & 16.7 & 3{,}162 \\
EfficientNet-B0~\cite{tan2019efficientnet} & 5.3M & 128 & 314 & 11.3 & 1{,}276 \\
ViT-B/16~\cite{dosovitskiy2021vit} & 86.6M & 64 & 211 & 9.0 & 1{,}910 \\
GPT-2~\cite{radford2019gpt2} & 124.4M & 8$\times$1024 & 275 & 21.2 & 2{,}115 \\
\bottomrule
\end{tabular}
\end{table}

\para{Tensor capture.} We intercept every ATen operator of a training iteration in PyTorch~\cite{paszke2019pytorch} with a dispatch-mode hook and, like object-level GPU memory profilers~\cite{lin2023drgpum}, record for every tensor storage its producing operator, its size, and every read and write. A tensor produced in the forward pass and read in the backward pass is a saved activation, exactly the set that rematerialization can evict; PyTorch's saved-tensor hooks confirm the set. We capture tensor contents on the CPU at a small batch (8 images, or one 1{,}024-token sequence) over three consecutive iterations with optimizer updates. We record operator traces at the training batch sizes on PyTorch's meta device, which propagates shapes without data. Each captured tensor maps to its trace counterpart by producing operator and output slot; GPT-2's operator sequence differs slightly between the two batch sizes, so we align the two sequences and treat the 4\% of its saved bytes left unmatched as incompressible. Two adjustments make the CPU run match GPU training: dropout runs through the fused kernel that GPUs use, which saves a 1-byte mask, and Transformer attention uses the frameworks' non-fused implementation, which saves the softmax output (Section~\ref{sec:discussion} discusses fused attention). FP32 tensors are compressed as captured. BF16 tensors are their round-to-nearest-even conversions, which approximates BF16 mixed-precision training~\cite{kalamkar2019bf16}. Tensors above 32\,MB are compressed on every fourth 4\,KB chunk.

\para{Simulator.} An event-driven simulator replays each trace on the platforms of Table~\ref{tab:platforms}. An operator takes a 4\,$\mu$s launch overhead plus the larger of its FLOPs over the achievable math throughput and its bytes over the achievable memory bandwidth. The simulator models one compute stream, a device-to-host and a host-to-device copy queue (FIFO, with 5\,$\mu$s setup per transfer), the compression engine, and a hard budget for activation memory. An allocation that does not fit waits for pending swap-outs to finish, and a prefetch starts at its scheduled point once the swap-out of the same tensor has finished. A plan whose replay overflows the budget is re-planned with a tighter target. On small synthetic traces, the simulator reproduces hand-computed schedules, including stalls behind slow swap-outs and late prefetches. The budget is a fraction of the peak activation memory without eviction; parameters, gradients, and optimizer state lie outside it, and the optimizer step is modeled as a memory-bound pass. The modeled ResNet-50 throughput on platform B is 2{,}529 images/s, within 3\% of the 2{,}470 images/s that NVIDIA reports for an A100 with mixed precision~\cite{nvidia2023dleresnet}.

\begin{table}[t]
\centering
\caption{Modeled platforms. Percentages give the achievable fraction of the peak; link and engine bandwidths are per direction.}
\label{tab:platforms}
\setlength{\tabcolsep}{4pt}
\footnotesize
\begin{tabular}{lll}
\toprule
 & Platform A & Platform B \\
\midrule
GPU & RTX 2080 Ti~\cite{nvidia2018turing} & A100 40\,GB~\cite{nvidia2020a100} \\
Precision & FP32 & BF16 mixed \\
Peak math & 13.4\,TFLOPS (70\%) & 312\,TFLOPS (60\%) \\
Memory bandwidth & 616\,GB/s (80\%) & 1{,}555\,GB/s (80\%) \\
Host link & PCIe 3.0, 12\,GB/s & PCIe 4.0, 24\,GB/s \\
Compression engine & 64\,GB/s & 64\,GB/s \\
\bottomrule
\end{tabular}
\end{table}

\para{Policies.} We compare six policies with the same simulator. The compressed policies plan with the ratios observed in one captured iteration and transfer with the ratios of the next.
\begin{itemize}
\item \textbf{RECOMP}: recomputation only, with DTR's heuristic~\cite{kirisame2021dtr}.
\item \textbf{SWAP}: swapping only, choosing tensors by the largest product of size and idle time, in the spirit of vDNN~\cite{rhu2016vdnn}.
\item \textbf{DTE}: MegTaiChi's cost model (Equation~\ref{eq:dte})~\cite{hu2022megtaichi}. We omit its fragmentation and recomputation-count terms, which address problems orthogonal to the choice between swapping and recomputation, and give every policy a fragmentation-free allocator, which MegTaiChi's TMA provides.
\item \textbf{BW}: the planner of Section~\ref{sec:planner} without compression, a Capuchin-style bandwidth-aware hybrid~\cite{peng2020capuchin}.
\item \textbf{BW+ZVC}: the same planner with the ZVC codec of cDMA~\cite{rhu2018cdma}.
\item \textbf{\sys{}}: the same planner with the \sys{} codec.
\end{itemize}

%% file: sections/eval.tex
\section{Evaluation}
\label{sec:eval}

\subsection{Compression Ratio}
\label{sec:eval-cr}

\begin{figure}[t]
\centering
\includegraphics[width=\columnwidth]{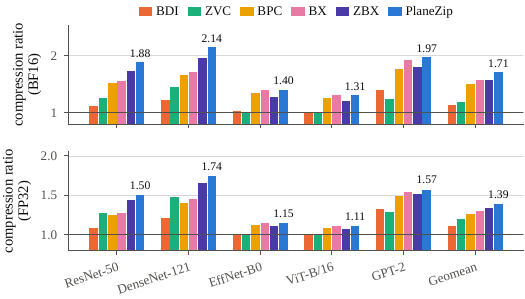}
\caption{Compression ratio of all saved activations per model and codec (higher is better). Labels give the \sys{} ratio.}
\label{fig:cr}
\end{figure}

Figure~\ref{fig:cr} compares six codecs on all saved activations of one iteration, weighted by bytes. In BF16, \sys{} compresses activations by 1.71$\times$ (geometric mean), from 1.31$\times$ for ViT-B/16 to 2.14$\times$ for DenseNet-121. ZVC compresses only the models with ReLU activations or causal masks and leaves EfficientNet-B0 and ViT-B/16 at 1.00$\times$, for a geometric mean of 1.18$\times$. BPC, which applies its delta transform to every block, reaches 1.50$\times$, and no single encoding exceeds 1.57$\times$: choosing per block adds 9\% over the best of them. FP32 activations compress less, 1.39$\times$ with \sys{}, because their 23-bit mantissas are nearly random and dilute the redundancy of the sign and exponent bits.

\input{figs/tab_classes.tex}

Table~\ref{tab:classes} breaks the ratios down by the operator that produced each tensor. Dense floating-point tensors (the outputs of convolution, linear, normalization, and elementwise operators) compress by 1.31--1.45$\times$, and BX encodes 74--98\% of their blocks: neighboring values share their high-order exponent bits but not their low-order bits. ReLU outputs compress by 3.21$\times$, against 2.16$\times$ with ZVC, and ZBX encodes 60\% of their blocks because it removes the zeros and then compresses the exponents of the survivors. The same mechanism lifts GPT-2's dropout outputs to 2.50$\times$ (ZVC: 1.94$\times$). Attention probabilities compress by 2.47$\times$ in GPT-2, whose causal mask zeroes half of them, and by 1.40$\times$ in ViT-B/16. One-byte dropout masks compress by 5.86$\times$ and the 64-bit indices of max pooling by 5.25$\times$. Coding BF16 blocks with 32-bit symbols, as the original BPC does, lowers the \sys{} ratio from 1.71$\times$ to 1.65$\times$: a 32-bit symbol is zero only if both of its BF16 values are.

\subsection{Predictability}
\label{sec:eval-pred}

\input{figs/tab_pred.tex}

Table~\ref{tab:pred} shows that compressibility is a stable property of each tensor. The ratio of each model changes by at most 1\% across three iterations with optimizer updates, and predicting each tensor's wire size from the previous iteration errs by 0.2--0.8\% of the wire bytes. A signature that keeps only the operator type, which would let one entry serve every layer, errs by 2.6--12.5\%, and a single global ratio by up to 43\%, because tensors of different layers and operators compress differently. Training progress matters more than iterations: at random initialization, the ratios differ from those of the trained models by $-$6\% (ResNet-50) to $+$12\% (GPT-2), so the runtime refreshes its table during training instead of reusing ratios across runs.

\subsection{Throughput Under a Memory Budget}
\label{sec:eval-main}

\begin{figure*}[t]
\centering
\includegraphics[width=\textwidth]{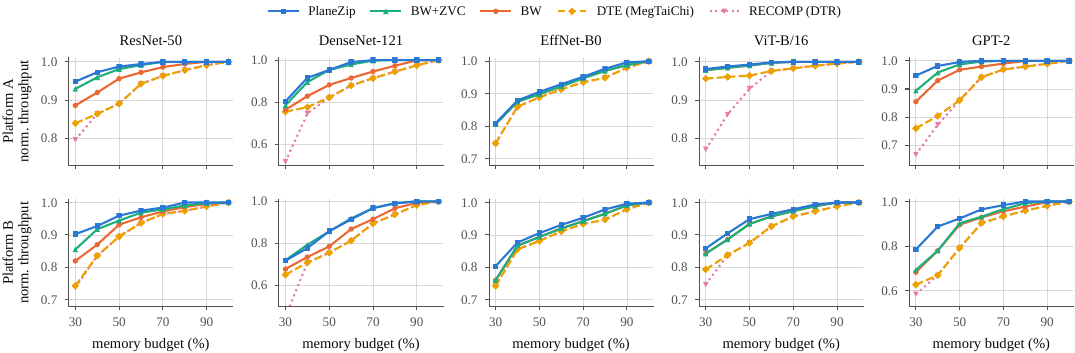}
\caption{Training throughput normalized to unlimited memory, as the activation-memory budget shrinks from 100\% to 30\% of the peak. SWAP is omitted because it falls below 0.5 at most budgets; RECOMP leaves the axis for DenseNet-121 at 30\%.}
\label{fig:main}
\end{figure*}

Figure~\ref{fig:main} shows the throughput of each policy as the budget shrinks. At 50\% of the peak, the mean overhead on platform A is 14.1\% for RECOMP, 13.3\% for DTE, 6.7\% for BW, 4.1\% for BW+ZVC, and 3.5\% for \sys{}; on platform B it is 19.6\%, 19.6\%, 13.0\%, 10.5\%, and 9.0\%. SWAP alone slows training by 2.8$\times$ (A) and 5.1$\times$ (B), because the link cannot move half of the activations within an iteration. DTE matches RECOMP in 58 of the 70 configurations: its cost model charges every swap its full transfer time and therefore picks recomputation. Bandwidth-aware swapping halves DTE's overhead on platform A and cuts it by a third on platform B, and \sys{} then removes another 47\% (A) and 31\% (B) of what remains. At 30\% of the peak, \sys{} cuts BW's overhead from 17.4\% to 12.2\% (A) and from 33.2\% to 23.7\% (B). Viewed the other way, \sys{} stays within 5\% of full speed down to a budget of 43\% of the peak on platform A and 59\% on platform B on average, against 54\% and 67\% for BW and 64\% and 75\% for DTE.

The gain grows with compressibility and is largest when the link can cover part, but not all, of the memory deficit. GPT-2 gains the most: its attention tensors, half of its saved bytes, compress by 2.8$\times$, so at 30\% on platform B its overhead falls from 46\% with BW to 28\%. DenseNet-121 gains 11 points at 50\% on platform B (27\% to 17\%), but BW+ZVC captures nearly all of it: the tensors worth swapping are sparse ReLU outputs that ZVC already compresses by 3.1$\times$, beyond the 2.7$\times$ at which the engine instead of the link limits the transfer. EfficientNet-B0 gains the least, 1.5 points at 50\% on platform B: its activations compress by only 1.40$\times$, and even compressed the link can move just 8\% of them. On platform A, ViT-B/16 computes slowly enough that BW hides most transfers (1.0\% overhead at 50\%), which leaves little for compression.

\subsection{Where the Gains Come From}
\label{sec:eval-why}

\begin{figure}[t]
\centering
\includegraphics[width=\columnwidth]{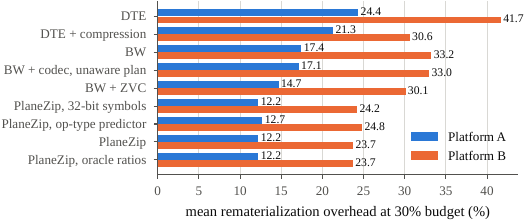}
\caption{Ablation at a budget of 30\% of the peak: mean overhead over the five models (lower is better).}
\label{fig:ablation}
\end{figure}

Figure~\ref{fig:ablation} isolates the components at a 30\% budget. Compressing the swaps of a compression-unaware plan recovers only 0.3 of the 5.2 points that separate BW from \sys{} on platform A and 0.2 of 9.5 points on platform B: shorter transfers free link time, but the plan does not fill it. In 67 of the 70 configurations, the planner keeps a compression-aware plan. Adding compression to DTE's cost model lowers its overhead from 24.4\% to 21.3\% (A) and from 41.7\% to 30.6\% (B), more than half of it through GPT-2, whose compressed attention probabilities become cheaper to swap than to recompute, but it still trails \sys{} by 9.1 and 6.9 points. With ZVC instead of the \sys{} codec, the planner keeps 52\% of the gain on platform A and 33\% on platform B, where dense BF16 tensors that ZVC cannot compress dominate. Coding with 32-bit symbols costs 0.5 points on platform B; on platform A, the FP32 symbols are 32 bits anyway. The predictor matters little as long as it is per tensor: oracle ratios leave the overhead unchanged, while the operator-type signature raises it by 0.5 (A) and 1.1 (B) points. Compared with BW, \sys{} swaps 1.5$\times$ (A) and 2.1$\times$ (B) more bytes and spends 30\% (A) and 31\% (B) less time on recomputation, while stalls stay at or below 1.1\% of the iteration.

\subsection{Sensitivity}
\label{sec:eval-sens}

\begin{figure}[t]
\centering
\includegraphics[width=\columnwidth]{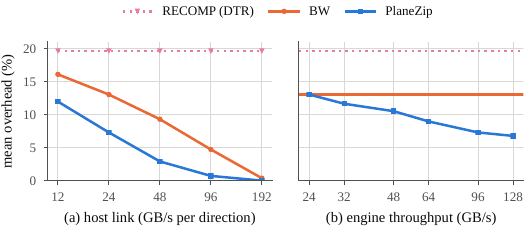}
\caption{Mean overhead at a 50\% budget on platform B. (a) Host-link bandwidth, with an engine four times faster than the link. (b) Engine throughput with a 24\,GB/s link.}
\label{fig:sens}
\end{figure}

Figure~\ref{fig:sens}a varies the link bandwidth of platform B from 12 to 192\,GB/s per direction. The gain peaks at intermediate bandwidths. At 12\,GB/s, which also approximates two PCIe~4.0 GPUs sharing a switch uplink, \sys{} lowers the mean overhead from 16.1\% (BW) to 12.0\%; at 24 and 48\,GB/s it removes 5.7 and 6.4 points; at 192\,GB/s the uncompressed link already carries everything worth swapping, and both policies stay at or below 0.4\%. Figure~\ref{fig:sens}b varies the engine throughput. An engine that is only as fast as the link never helps, because compressed transfers become engine-bound and the runtime bypasses it. At the default 64\,GB/s the overhead is 9.0\%, and the 96\,GB/s of the two lanes estimated in Section~\ref{sec:hw} lowers it to 7.3\%.

\subsection{Discussion and Limitations}
\label{sec:discussion}

\para{Fused attention.} FlashAttention~\cite{dao2022flashattention} recomputes the attention matrix instead of saving it. In GPT-2, the attention probabilities, their dropout outputs, and their masks account for 51\% of the saved bytes and compress by 2.82$\times$, helped by the zeros of the causal mask; the remaining tensors compress by 1.46$\times$, like the dense tensors of the other models. With fused attention, GPT-2 would gain less, closer to ViT-B/16 in Figure~\ref{fig:main}.

\para{Fidelity.} Our timing model ignores kernel fusion and interference between copies and kernels, our ratios come from CPU-executed iterations at small batch sizes, and our BF16 tensors are conversions. Capturing tensors in GPU runs with an accelerator analysis framework such as PASTA~\cite{lin2026pasta} would remove the last two approximations. The timing model matches published ResNet-50 throughput within 3\%, the ratios reflect real data, weights, and training updates, and all policies share the same model, but absolute overheads on real hardware will differ. The planner also assumes that the operator sequence repeats; models with dynamic control flow need one plan per sequence, or MegTaiChi's online eviction with the compressed cost of Equation~\ref{eq:dte}.

\para{Software compression.} GPU kernels could replace the engine, with a warp transposing a 128-byte block into bit-planes in 32 or 64 ballot instructions, but they would compete with training for SMs and memory bandwidth.

\para{Other host traffic.} The engine could compress any data that crosses the link, such as UVM page migrations in oversubscribed training~\cite{lin2025understanding} and LLM inference~\cite{lin2026autouvm}, or the data that CPU-GPU hybrid LLM inference exchanges~\cite{lin2026hybridgen}.

%% file: figs/tab_classes.tex
\begin{table}[t]
\centering
\caption{Saved activations by producing operator (BF16). Share: range of the per-model shares of saved bytes. Ratios: byte-weighted over all models, with the 3-bit tag included for \sys{}. Top: most frequent encoder.}
\label{tab:classes}
\setlength{\tabcolsep}{3.4pt}
\footnotesize
\begin{tabular}{lrrrrl}
\toprule
Producer & Share (\%) & ZVC & BX & \sys{} & Top \\
\midrule
Conv output & 18--52 & 1.00 & 1.43 & \textbf{1.45} & BX (74\%) \\
Elementwise & 26--39 & 1.00 & 1.33 & \textbf{1.33} & BX (88\%) \\
ReLU output & 45--48 & 2.16 & 1.85 & \textbf{3.21} & ZBX (60\%) \\
Copy/concat & 6--32 & 1.02 & 1.41 & \textbf{1.41} & BX (84\%) \\
Softmax & 16--23 & 1.51 & 2.02 & \textbf{2.10} & BX (62\%) \\
Linear output & 6--21 & 1.00 & 1.32 & \textbf{1.32} & BX (96\%) \\
Dropout output & 23 & 1.94 & 2.21 & \textbf{2.50} & ZBX (51\%) \\
Norm in/out & 0--11 & 1.00 & 1.31 & \textbf{1.31} & BX (98\%) \\
Log-softmax & 0--8 & 1.00 & 2.20 & \textbf{2.23} & DBX (72\%) \\
Dropout mask & 6 & 1.00 & 5.96 & \textbf{5.86} & BX (100\%) \\
Indices & 0--2 & 1.88 & 3.01 & \textbf{5.25} & ZBX (100\%) \\
\bottomrule
\end{tabular}
\end{table}

%% file: figs/tab_pred.tex
\begin{table}[t]
\centering
\caption{Stability of compressibility (\sys{} codec, BF16): ratio in three consecutive iterations, wire-size error of three predictor signatures (absolute errors over total wire bytes), and ratio at random initialization.}
\label{tab:pred}
\setlength{\tabcolsep}{3pt}
\footnotesize
\begin{tabular}{lccccc}
\toprule
 & Ratio per & \multicolumn{3}{c}{Wire-size error (\%)} & Random \\
\cmidrule(lr){3-5}
Model & iteration & Tensor & Op type & Global & init. \\
\midrule
ResNet-50 & 1.88/1.88/1.89 & 0.6 & 9.7 & 28.7 & 1.78 \\
DenseNet-121 & 2.14/2.13/2.15 & 0.7 & 12.5 & 43.1 & -- \\
EffNet-B0 & 1.40/1.40/1.41 & 0.8 & 4.3 & 4.8 & -- \\
ViT-B/16 & 1.31/1.31/1.31 & 0.3 & 2.6 & 4.3 & 1.41 \\
GPT-2 & 1.97/1.97/1.97 & 0.2 & 3.2 & 29.8 & 2.22 \\
\bottomrule
\end{tabular}
\end{table}

%% file: sections/related.tex
\section{Related Work}
\label{sec:related}

\para{Rematerialization.} Checkpointing~\cite{chen2016sublinear} and its optimal ILP formulation~\cite{jain2020checkmate} trade memory for recomputation statically; DTR~\cite{kirisame2021dtr} does so online with a greedy heuristic. vDNN~\cite{rhu2016vdnn} and SwapAdvisor~\cite{huang2020swapadvisor} swap activations to host memory; Capuchin~\cite{peng2020capuchin} and MegTaiChi~\cite{hu2022megtaichi} combine swapping and recomputation, and MegTaiChi adds tensor partitioning and a fragmentation-free allocator. These planners treat a tensor's transfer size as its raw size; \sys{} supplies compressed sizes and a compression engine, and its planner could be replaced by any of them that accepts per-tensor transfer costs. Profilers such as DrGPUM~\cite{lin2023drgpum} remove memory that an application wastes; rematerialization handles the memory it needs.

\para{Activation compression.} cDMA~\cite{rhu2018cdma} is the closest work: a ZVC compressor in the DMA engine that accelerates vDNN's swaps. \sys{} extends the codec beyond zeros, which matters for activations without ReLU sparsity (Table~\ref{tab:classes}), and makes the planner aware of the compression, without which the gain largely disappears (Section~\ref{sec:eval-why}). Gist~\cite{jain2018gist} encodes ReLU outputs as bitmaps or sparse matrices and stores other activations in reduced precision, and ActNN~\cite{chen2021actnn} quantizes activations to 2 bits. Lossy schemes reach higher ratios but change the numerics of training; \sys{} is lossless.

\para{Hardware memory compression.} BDI~\cite{pekhimenko2012bdi} and BPC~\cite{kim2016bpc} compress cache lines and memory blocks; Compresso~\cite{choukse2018compresso} manages compressed main memory, and Buddy Compression~\cite{choukse2020buddy} applies BPC to GPU memory and spills incompressible lines to a slower peer memory. \sys{} reuses BPC's transform and code map and Compresso's chunked allocation, but compresses only the data that crosses the slow link, where the latency of compression is hidden. ZipNN~\cite{hershcovitch2024zipnn} exploits the same exponent redundancy in stored model weights. Baryon~\cite{li2023baryon} combines compression and sub-blocking in hybrid memory, and \sys{} carries the predictor-splitting idea of Yokoyama et al.~\cite{yokoyama2026rvv} over to compressibility.

%% file: sections/conclusion.tex
\section{Conclusion}
\label{sec:conclusion}

GPU compute has outgrown the host link, so rematerialization planners recompute almost everything and swap only what fits in idle link time. \sys{} enlarges that budget by compressing swapped activations with a per-block choice among four bit-plane and zero-value encodings, and turns the compression into speed by planning swaps in compressed bytes. Activations from five models compress by 1.71$\times$ in BF16 and 1.39$\times$ in FP32, each tensor's compressibility carries over from one iteration to the next, and at half the activation memory the rematerialization overhead falls by 47\% and 31\% on the two platforms relative to a bandwidth-aware planner without compression. Next steps are a prototype on a GPU copy engine or in software, and compressed in-place storage for tensors that are expensive to recompute.